\documentclass{aa}  
\usepackage{graphicx}
\usepackage{txfonts}
\usepackage{natbib}
\usepackage{dcolumn}
\usepackage{placeins}

\newcolumntype{.}{D{.}{.}{-1}}
\newcolumntype{;}{D{;}{.}{7}}
\bibpunct{(}{)}{;}{a}{}{,} 

\begin{document}
   \title{Molecular gas in nearby low-luminosity QSO host galaxies}


    \author{T. Bertram \inst{1}
    \and A. Eckart \inst{1} 
    \and S. Fischer \inst{1}
    \and J. Zuther \inst{1}
    \and C. Straubmeier \inst{1}
    \and L. Wisotzki \inst{2}
    \and M. Krips \inst{3}
    }
  
   \offprints{T. Bertram}

\institute{I. Physikalisches Institut, Universit\"at zu K\"oln,
           Z\"ulpicher Str. 77, 50937 K\"oln, Germany\\
	   \email{bertram@ph1.uni-koeln.de}
	   \and Astrophysikalisches Institut Potsdam, An der Sternwarte 16, 14482 Potsdam
	   \and Harvard-Smithsonian Center for Astrophysics, SMA project, 645 North A‘ohoku Place, Hilo, HI 96720
          }

   \date{Received 2 April 2007 / Accepted 14 May 2007}

 
  \abstract
   {}
   {This paper addresses the global molecular gas properties of a representative sample of galaxies hosting low-luminosity quasistellar objects. An abundant supply of gas is necessary to fuel both the active galactic nucleus and any  circum-nuclear starburst activity of QSOs. The connection between ultraluminous infrared galaxies and the host properties of QSOs is still subject to a controversial debate. 
   Nearby low-luminosity QSOs are ideally suited to study the properties of their host galaxies because of their higher frequency of occurrence compared to high-luminosity QSOs in the same comoving volume and because of their small cosmological distance. }
   { We selected a  sample of nearby low-luminosity QSO host galaxies that is free of infrared excess biases. All objects are drawn from the Hamburg-ESO survey for bright UV-excess QSOs, have $\delta$$>$-30$^\circ$ and redshifts that do not exceed z=0.06.  The IRAM 30m telescope was used to measure the \element[][12]{CO}(1$-$0) and \element[][12]{CO}(2$-$1) transition in parallel.}
   {27 out of 39 galaxies in the  sample have been detected. The molecular gas masses of the detected sources range from $\rm 0.4\cdot10^9M_\odot$ to $\rm 9.7\cdot10^9M_\odot$. The upper limits of the non-detected sources correspond to molecular gas masses between  $\bf \rm 0.3\cdot10^9M_\odot$ and $\bf \rm 1.2\cdot10^9M_\odot$. We can confirm that the majority of galaxies hosting low-luminosity QSOs are rich in molecular gas.  
   The properties of galaxies hosting brighter type I AGN and circumnuclear starformation regions differ from the properties of galaxies with fainter central regions. The overall supply of molecular gas and the spread of the linewidth distribution is larger.       
   When comparing the far-infrared with the CO luminosities, the distribution can be separated into two different power-laws: one describing the lower activity Seyfert I population and the second describing the luminous QSO population.  The separation in the $L_{\rm FIR}$/$L'_{\rm CO}$ behaviour may be explainable with differing degrees of compactness of the emission regions. We provide a simple model to describe the two power-laws. The sample studied in this paper is located in a transition region between the two populations.
   } 
   {}


   \maketitle
\section{Introduction}
The investigation of the molecular gas content and its distribution in the host galaxies of quasistellar objects (QSOs) is a key issue in the understanding of evolutionary sequences and environments of active galactic nuclei (AGN).
\
All proposed sequences involve star formation and an abundant supply of material to fuel the starburst and the central engine. 

\citet{1988apj...325...74s} were the first to discuss a connection between ultraluminous infrared galaxies (ULIGs) and QSOs, suggesting that both represent different stages in an evolutionary sequence that starts with the collision of gas-rich spiral galaxies. In the course of the merger, circumnuclear starburst activity is triggered and a dominant part of available molecular gas is concentrated within the center of the merging galaxy.   
\
Several interferometric studies on individual ULIGs, like the archetypal ULIG Arp~220 \citep[e.g.][]{1991apj...366l...5s},  reveal a massive concentration of molecular gas in the center ($\la$1 kpc diameter). The bulk of H$_2$ in these gas rich objects seems to be associated with this central gas accumulation. This massive concentration not only favors starburst activity but may also be required to drive the nuclear activity that becomes apparent in the QSO phase of the sequence. The more distant and more gas-rich submillimeter galaxies (SMGs) resemble scaled-up versions of the local ULIG population, also showing compact CO emission regions confined to the center. \citet{2006apj...640..228t} state a median diameter of $\leq$4~kpc for 8 SMGs. 

Contrary to the evolutionary model, \citet{1998apj...507..615d} question the need for an AGN to power the FIR emission in ULIGs. They argue that extreme starbursts in circumnuclear molecular disks or rings are fully accountable for the high FIR luminosity. Following their line of arguments implies that, although several ULIGs with AGN are known, AGN are not mandatory for the ULIG appearance and ULIGs are not necessarily the predecessors of QSOs.   

A recent study of \citet{2003mnras.340.1095d} on the morphological properties of QSO host galaxies discusses a decrease of the number density of disk dominated host galaxies ending in a complete depletion at $M_{\rm V}\le-23$ (assuming H$_0=50$ km s$^{-1}$ Mpc$^{-1}$). From their analysis of Hubble space telescope and near infrared (NIR) data they conclude that QSO hosts have properties  very similar to red, quiescent elliptical galaxies. These objects usually are gas depleted.   
\
\citet{2006a&a...458..107b} argues for a more differentiated picture at least of intermediate luminosity QSO hosts and points to several indications of recent or ongoing star formation activity in these objecs.  

Little is known about molecular gas in nearby QSO hosts: \citet{2001aj....121.3285e,2006aj....132.2398e} and \citet{2003apj...585l.105s} discuss  small samples of QSO host galaxies selected from the Palomar-Green (PG) Bright Quasar Survey \citep{1983apj...269..352s}. Few studies on individual sources add to the incomplete picture of molecular gas in local QSO hosts.  
A common denominator seems to be the presence of large amounts of molecular gas in the majority of these objects, which militates against a quiescent nature of host galaxies. The distribution of molecular gas, the extent and density of regions emitting CO line emission and similarities with or differences to local Seyfert I galaxies have not yet been studied on a solid statistical base.  
\
More detailed multi-wavelength investigations of a larger sample of nearby QSOs are certainly beneficial in the controversial debate on the nature and history of QSOs.   
\
Especially the separation of the starburst and the AGN component in extragalactic objects from the faint contribution of the underlying host galaxy  require exceptionally high spatial resolution and sensitivity. 
\
The nearby QSOs with z$\le$0.1, therefore, represent an important link between the cosmologically local, less luminous AGN and the high redshift, high luminosity QSOs (at z$\ge$0.5).
\
\begin{figure}
  \centering
  \resizebox{\hsize}{!}{\includegraphics[]{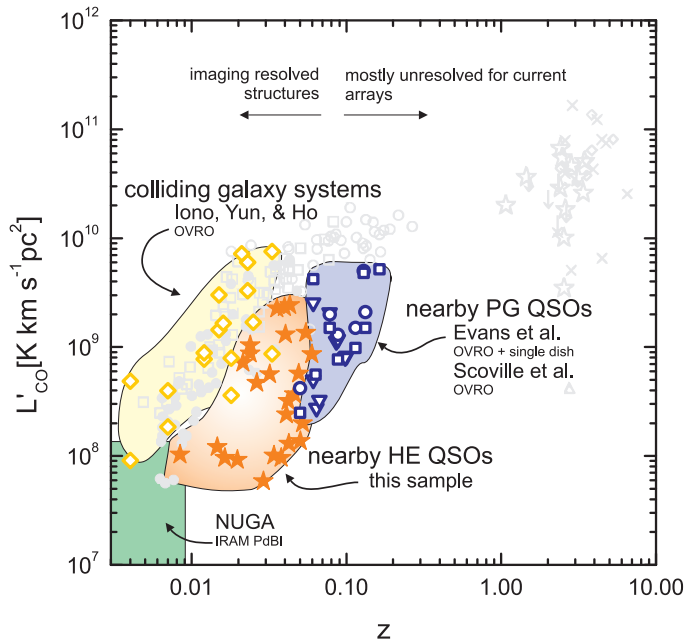}} 
  \caption{$L_{\rm CO}$-z distribution of already existing interferometric imaging studies of AGN hosts / interacting galaxies (thick outlined symbols) and the distribution of the nearby low-luminosity QSO sample. The underlying light gray symbols represent a compilation of the most CO luminous objects ((U)LIGs/SMGs/HzRGs/QSOs) measured so far \citep{2005mnras.359.1165g} for the corresponding redshifts. The objects in the nearby low-luminosity QSO sample have redshifts  that allow to resolve structures on the sub-kpc scale in the \element[][12]{CO}(1$-$0) transition with current mm-interferometers. \emph{[See the online edition of the Journal for a color version of this figure.]}}   
  \label{FigCOsurveys}
\end{figure}
Within this volume the accessibility of important structural information with state-of-the-art interferometers is feasible. In the millimeter domain, interferometers then provide sub-kpc resolutions that allow to probe the sizes of potentially compact circum-nuclear molecular gas reservoirs similar to the ULIG case (cf. Fig. \ref{FigCOsurveys}). If the molecular gas is less confined to the center compared to the ULIG case it will be possible to study the distribution and potential signs of interaction on larger scales.   

We have selected a sample of nearby UV excess QSOs or luminous Seyfert I galaxies, that are not only suitable for interferometric imaging in the millimeter wavelength domain but also allow for detailed imaging and spectroscopy in the NIR. For 41 members of the sample single-dish CO data was obtained (in one case also interferometric data). The sample is introduced in Sect. \ref{SectQSOSample}, together with a definition of the term ``low-luminosity QSO'' that is used throughout this paper.  
\
The results of the observations follow in Sect. \ref{SectResults}. The data allows to draw conclusions on the total molecular gas content of low-luminosity QSOs and can be related to existing visible and far infrared quantities, as presented in Sect. \ref{SectDiscussion}. H$_0$=75 km s$^{-1}$ Mpc$^{-1}$ and q$_0$=0.5 are assumed throughout the paper.

\section{The nearby low-luminosity QSO sample}
\label{SectQSOSample}
The only selection criterion for the sample of nearby low-luminosity QSOs was their small cosmological distance:  only objects with a redshift z$<$0.060 were chosen. 
\
This redshift limit is based on a NIR spectroscopic constraint: it ensures the observability of the diagnostic  CO(2$-$0) rotation vibrational band head absorption line, which is important for the stellar population analysis. This line is then still accessible in the K-band.    

The members were selected entirely from an extended catalog of sources found in the Hamburg/ESO survey (HES). The HES \citep{2000a&a...358...77w} is a wide angle survey for optically bright QSOs, with a well-defined flux limit of $B_J \la 17.3$, varying from field to field, and a redshift coverage of 0$<$z$<$3.2.  QSO candidates were identified in digitized objective prism plate data by applying a color- and spectral feature based selection scheme. With the exception of objects listed in the \citet{1996cqan.book.....v} catalog, all candidates were subject to spectroscopic follow-up observations to confirm the object's identity and to reject false classifications. 

The application of a starlike morphological criterion in many other QSO samples like the PG Bright Quasar Survey or the 10k catalog of the 2dF QSO Redshift Survey results in a significant degree of incompleteness at the low redshift end. One of the main advantages of the HES over other surveys is the consideration of extended objects. The HES sample, therefore, shows a high volume density of luminous type I AGN also at low cosmological distances. This circumstance makes the HES sample a valuable source of objects for a study of nearby QSOs. A total of 99 objects within the volume 0.01$\le$z$\le$0.06 were identified by the HES. Thirty-nine of these form the  subsample that is discussed in this paper. The objects in the subsample all have $\delta$$>$-30$^\circ$. To avoid potential selection biases, the composition of the subsample matches the full sample of nearby low-luminosity QSOs in terms of redshift distribution and percentage of IRAS detected sources.

It is important to note that no luminosity discrimination between QSOs and
Seyfert I galaxies was applied by the HES -- all luminous type I AGN showing broad emission lines (FWHM$\geq$1000 km s$^{-1}$) in their follow-up spectra were included in their catalog. This has a direct implication on the absolute brightness distribution of the subsample here referred to as ``nearby low-luminosity QSO sample''. Our sample clearly probes the low luminosity tail of the local quasar luminosity function \citep{1997a&a...325..502k}. 
All objects in the sample have absolute $B_{\rm J}$ magnitudes exceeding (i.e., dimmer than) the traditional boundary\footnote{translated to the cosmology used throughout the paper} $M_{\rm B}\sim$-22  between higher luminosity QSOs and lower luminosity Seyfert I galaxies. This boundary has no astrophysical motivation, as it was technologically induced at the time of its introduction \citep{1983apj...269..352s}. However, to respect the commonly used definition of the term ``QSO'', we explicitely use the term ``low-luminosity QSO'' throughout the paper for objects  identified in QSO surveys that may be fainter than the traditional boundary magnitude. Not only the HES but also the PG Bright Quasar Survey provide low-luminosity QSOs in their samples.   

The restriction to type I AGN in the HES naturally resulted in the exclusion of type II AGN in the nearby low-luminosity QSO sample. For this reason the comparison with literature data in Sect. \ref{SectDiscussion} was also restricted to type I AGN.   

Several members of the nearby low-luminosity QSO sample were already subject to studies in the NIR and mm wavelength domain. First NIR imaging and spectroscopic results of nearby low-luminosity QSO sample members are presented in \citet{2006a&a...452..827f}.
A series of observations carried out with SEST and BIMA \citep{2006newar..50..712b} preceded the observations presented in this paper.  In these first runs a slightly differently composed sample was scanned for millimetric CO emission. The primary goal was to identify the CO brightest objects for high resolution interferometric follow-up observations. The resulting SEST spectra and CO properties of two detected nearby HES objects, \object{HE~0108-4743} and \object{HE~2211-3903}, are included in Sections \ref{SectResults} and \ref{SectDiscussion}. However, due to the higher detection limit, they were not included in the FIR unbiased subsample discussed in Sect. \ref{SectDetect}. The third detection, \object{HE~1029-1831}, was included. Follow-up observations with the Plateau de Bure Interferometer allowed to resolve the central region in this exemplary case. A detailed analysis can be found in \citet{2007a&a...464..187k}.


\section{Observations and data reduction}

\begin{table*}
\begin{minipage}[t]{\textwidth}
\begin{center}
\caption{Journal of the observations carried out with the IRAM 30m telescope in in September 2005, June 2006, and January 2007. Three previously detected sources are also listed. z was taken from the HES. $t_{\rm int}$ denotes the total on-source integration time for each receiver.}
\label{TabSources}

\renewcommand{\footnoterule}{}  
\begin{tabular*}{\linewidth}{@{\extracolsep\fill}llp{0.3cm}rrrp{0.3cm}rrrp{0.5cm}.lr}
\hline
\hline\\
\multicolumn{1}{c}{Obj.} &
\multicolumn{1}{c}{alt. name} & &
\multicolumn{3}{c}{R.A. (J2000)} & &
\multicolumn{3}{c}{Dec. (J2000)} & &
\multicolumn{1}{c}{z} & 

\multicolumn{1}{c}{Obs.}&
\multicolumn{1}{c}{$t_{\rm int}$}\\
     & & &
\multicolumn{1}{r}{[ h} &
\multicolumn{1}{r}{m} &
\multicolumn{1}{r}{s]} & &
\multicolumn{1}{r}{[ $^{\circ}$} &
\multicolumn{1}{r}{\arcmin} &
\multicolumn{1}{r}{\arcsec]} & 
& &
\multicolumn{1}{c}{run}&
\multicolumn{1}{c}{[min]} \\
\hline\\
 
 \object{HE 0021-1810} & \object{VCV2001 J002339.2-175355} & & 00 & 23 & 39.4 & &-17 & 53 & 53 & & 0.053  &'07   & 118   \\
 \object{HE 0021-1819} & \object{NPM1G -18.0010}           & & 00 & 23 & 55.3 & &-18 & 02 & 50 & & 0.052  &'07   & 140   \\
 \object{HE 0040-1105} & \object{VIII Zw 36}               & & 00 & 42 & 36.8 & &-10 & 49 & 21 & & 0.041  &'06   & 112   \\        
 \object{HE 0045-2145} & \object{IRAS 00452-2145 }         & & 00 & 47 & 41.3 & &-21 & 29 & 27 & & 0.021  &'05   &  45   \\      
 \object{HE 0108-4743} & \object{IRAS 01089-4743 }         & & 01 & 11 & 09.7 & &-47 & 27 & 36 & & 0.029\footnotemark[1]  &SEST  &  76 \\      
 \object{HE 0114-0015} & \object{SDSS J011703.58+000027.4} & & 01 & 17 & 03.6 & &+00 & 00 & 27 & & 0.046  &'06   &  98 \\      
 \object{HE 0119-0118} & \object{II Zw  1}                 & & 01 & 21 & 59.8 & &-01 & 02 & 25 & & 0.054  &'05   &  60 \\      
 \object{HE 0150-0344} & \object{IRAS 01505-0343}          & & 01 & 53 & 01.4 & &-03 & 29 & 24 & & 0.046  &'06   &  70 \\      
 \object{HE 0203-0031} & \object{Mrk 1018}                 & & 02 &  6 & 15.9 & &-00 & 17 & 29 & & 0.043  &'06   &  56 \\      
 \object{HE 0212-0059} & \object{Mrk 590}                  & & 02 & 14 & 33.6 & &-00 & 46 & 00 & & 0.027  &'05   &  60 \\      
 \object{HE 0224-2834} & \object{AM 0224-283}              & & 02 & 26 & 25.7 & &-28 & 20 & 59 & & 0.060  &'06   &  55 \\      
 \object{HE 0227-0913} & \object{Mrk 1044}                 & & 02 & 30 & 05.4 & &-08 & 59 & 53 & & 0.017  &'06   &  46 \\      
 \object{HE 0232-0900} & \object{NGC  985}                 & & 02 & 34 & 37.7 & &-08 & 47 & 16 & & 0.043  &'06   &  32 \\      
 \object{HE 0253-1641} & \object{NPM1G-16.0109}            & & 02 & 56 & 02.6 & &-16 & 29 & 16 & & 0.032  &'05   & 120 \\      
 \object{HE 0345+0056} & \object{IRAS 03450+0055}          & & 03 & 47 & 40.2 & &+01 & 05 & 14 & & 0.029  &'05   & 117 \\      
 \object{HE 0351+0240} & \object{VCV2001 J035409.4+024931} & & 03 & 54 & 09.4 & &+02 & 49 & 30 & & 0.034  &'06   &  74 \\      
 \object{HE 0412-0803} & \object{MS 04124-0802}            & & 04 & 14 & 52.6 & &-07 & 55 & 41 & & 0.038  &'06   & 115 \\      
 \object{HE 0429-0247} & \object{RXS J04316-0241}          & & 04 & 31 & 37.0 & &-02 & 41 & 25 & & 0.041  &'06   &  61 \\      
 \object{HE 0433-1028} & \object{Mrk  618}                 & & 04 & 36 & 22.2 & &-10 & 22 & 33 & & 0.033  &'05   &  25 \\      
 \object{HE 0853-0126} &                                   & & 08 & 56 & 17.8 & &-01 & 38 & 07 & & 0.060  &'06   & 135 \\      
 \object{HE 0853+0102} &                                   & & 08 & 55 & 54.3 & &+00 & 51 & 10 & & 0.052  &'06   & 120 \\      
 \object{HE 0934+0119} & \object{Mrk 707}                  & & 09 & 37 & 01.0 & &+01 & 05 & 43 & & 0.051  &'06   &  72 \\      
 \object{HE 0949-0122} & \object{Mrk 1239}                 & & 09 & 52 & 18.9 & &-01 & 36 & 44 & & 0.019  &'05   & 110 \\      
 \object{HE 1011-0403} & \object{PG 1011-040}              & & 10 & 14 & 20.6 & &-04 & 18 & 41 & & 0.057  &'06   & 122 \\      
 \object{HE 1017-0305} & \object{Mrk 1253}                 & & 10 & 19 & 32.9 & &-03 & 20 & 15 & & 0.048  &'05   & 274 \\      
 \object{HE 1029-1831} & \object{NPM1G-18.0348}            & & 10 & 31 & 57.3 & &-18 & 46 & 34 & & 0.040  &PdBI  &     \\
 \object{HE 1107-0813} &                                   & & 11 & 09 & 48.5 & &-08 & 30 & 15 & & 0.057  &'07   &  80 \\
 \object{HE 1108-2813} & \object{VCV2001 J111048.0-283004} & & 11 & 10 & 48.0 & &-28 & 30 & 03 & & 0.023  &'05, '07 & 41\\
 \object{HE 1126-0407} & \object{Mrk 1298, PG1126-041}     & & 11 & 29 & 16.6 & &-04 & 24 & 08 & & 0.060  &'05, '07 & 41\\
 \object{HE 1237-0504} & \object{NGC 4593}                 & & 12 & 39 & 39.4 & &-05 & 20 & 40 & & 0.009  &'05   &  35 \\      
 \object{HE 1248-1356} &                                   & & 12 & 51 & 32.4 & &-14 & 13 & 17 & & 0.015  &'06   & 102 \\
 \object{HE 1310-1051} & \object{PG 1310-108}              & & 13 & 13 & 05.7 & &-11 & 07 & 42 & & 0.034  &'07   &  64 \\
 \object{HE 1330-1013} &                                   & & 13 & 32 & 39.1 & &-10 & 28 & 53 & & 0.022  &'07   &  59 \\
 \object{HE 1338-1423} &                                   & & 13 & 41 & 12.9 & &-14 & 38 & 40 & & 0.041  &'07   &  85 \\
 \object{HE 1353-1917} & \object{VCV2001 J135636.7-193144} & & 13 & 56 & 36.7 & &-19 & 31 & 44 & & 0.034  &'07   &  97 \\
 \object{HE 1417-0909} &                                   & & 14 & 20 & 06.3 & &-09 & 23 & 13 & & 0.044  &'07   & 106 \\
 \object{HE 2128-0221} & \object{6dF J2130499-020814}      & & 21 & 30 & 49.9 & &-02 & 08 & 15 & & 0.052  &'05, '07   & 295\\      
 \object{HE 2211-3903} &                                   & & 22 & 14 & 42.0 & &-38 & 48 & 24 & & 0.039  &SEST  & 114 \\      
 \object{HE 2222-0026} & \object{SDSS J222435.29-001103.8} & & 22 & 24 & 35.3 & &-00 & 11 & 04 & & 0.058  &'07   & 225 \\
 \object{HE 2233-0124} & \object{VCV2001 J223541.9+013933} & & 22 & 35 & 41.9 & &-01 & 39 & 33 & & 0.056  &'07   & 224 \\
 \object{HE 2302-0857} & \object{Mrk 926}                  & & 23 & 04 & 43.4 & &-08 & 41 & 09 & & 0.046  &'07   & 150 \\
\hline
\end{tabular*}\\
\end{center} 
\footnotemark[1] The redshift for \object{HE~0108-4743} was taken from \citet{1996cqan.book.....v} and not spectroscopically verified by the HES. Since  it differs significantly from other visual redshift data \citep{1999mnras.308..897l}, the value was excluded from the redshift discussion in Sect. \ref{SectRedshift}.  
\end{minipage}
\end{table*}
\.
In this paper we report on sensitive observations carried out with the IRAM 30m telescope on Pico Veleta (Spain) in September 2005, June 2006 and January 2007. 

Twenty-six out of 38 observed sources were detected (cf. Table~\ref{TabSources}).  In most cases the detection limits were much lower than in the preceding observing runs. However, part of the data obtained in June 2006 suffered from mediocre weather conditions and had to be excluded from further analysis. Moreover, in several cases the observing frequencies had to be based only on the redshifts provided by the HES. Although nominally uncertain by 300 km s$^{-1}$, in some cases the Hamburg/ESO based systemic velocities deviated from the measured systemic velocity centroids by up to 750 km s$^{-1}$ (cf. Sect. \ref{SectRedshift}). A certain fraction of weak sources may not have been detected because of this de facto uncertainty.

The A and B receivers were used together with the 1 MHz and 4 MHz filter banks to acquire \element[][12]{CO}(1$-$0) and \element[][12]{CO}(2$-$1) emission line data in parallel. The configuration of the backends allowed for a velocity coverage $\Delta v$ between 1300 km s$^{-1}$ and 1400 km s$^{-1}$ at the redshifted CO lines. The compactness of the sources allowed to take advantage of the wobbling secondary and the resulting excellent baselines. In the three observing runs useful data with a total of $\sim$65h on-source integration time was obtained.

The data reduction was carried out using the IRAM software CLASS/GILDAS. Each polarization direction was analyzed individually and both directions were averaged whenever appropriate. In all cases it was sufficient to subtract linear baselines. To increase the signal-to-noise ratio, most of the spectra (cf. Fig. \ref{FigSpectra}) were hanning smoothed.  In order to obtain main beam temperatures, the main beam efficiency $B_{\rm eff}=0.75$ and the forward efficiency $F_{\rm eff}=0.95$ were applied to $T_{\rm A}^*$ at $\sim$110 GHz and $B_{\rm eff}=0.54$ and $F_{\rm eff}=0.91$ to $T_{\rm A}^*$ at $\sim$220 GHz.  The line intensities $I_{\rm CO}$ represent main beam temperatures integrated over the full velocity range of the emission line. The errors $\Delta I_{\rm CO}$ were determined by the geometric average of the line error $\Delta I_{\rm L} = \sigma v_{\rm res} \sqrt{N_{\rm L}}$  and the baseline error $\Delta I_{\rm B} = \sigma v_{\rm res} N_{\rm L} /\sqrt{N_{\rm B}}$, where $\sigma$ is the channel-to-channel rms noise, $v_{\rm res}$ the spectral resolution,  $N_{\rm L}$ the number of channels over which the line is distributed and $N_{\rm B}$ the number of channels used for the baseline fit.
\
\begin{figure*}
  \centering
  \resizebox{\hsize}{!}{\includegraphics{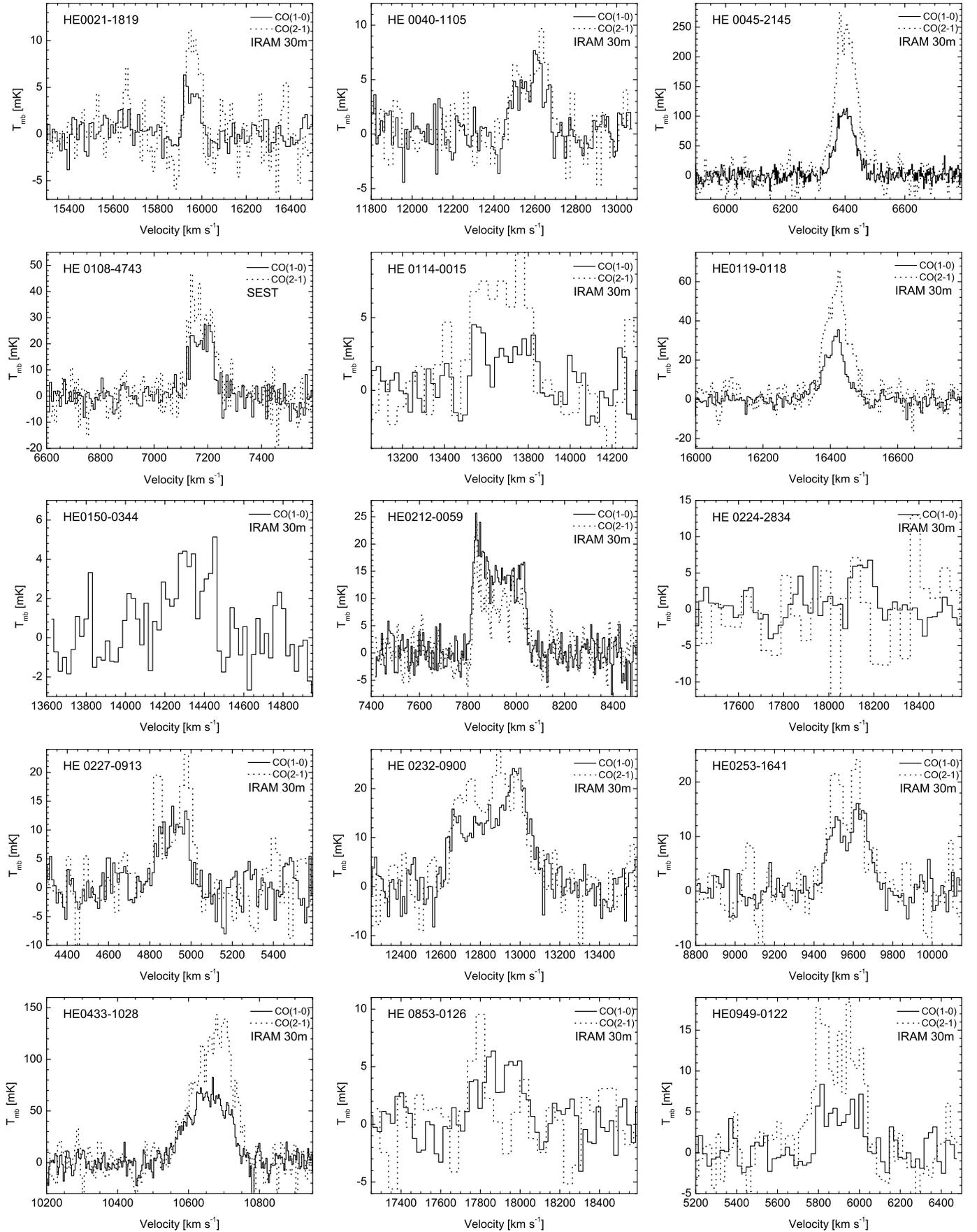}} 
  \caption{\element[][12]{CO} spectra of detected host galaxies. In two cases, the weather conditions prevented reliable measurements of the \element[][12]{CO}(2$-$1) transition.}
  \label{FigSpectra}
\end{figure*}
\
\addtocounter{figure}{-1}
\begin{figure*}
  \centering
  \resizebox{\hsize}{!}{\includegraphics{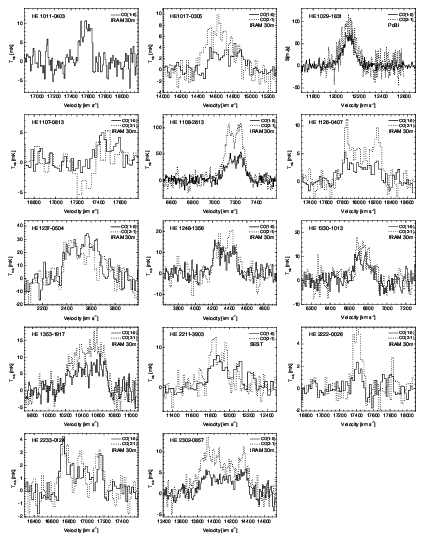}}
  \caption{---continued}   
\end{figure*}

\section{Results}
\label{SectResults}

\begin{table*}
\begin{minipage}[t]{\textwidth}
\begin{center}
\caption{Summary of \element[][12]{CO} properties. Unless otherwise noted, the data was acquired with the IRAM  30m telescope.}
\label{TabResults}

\renewcommand{\footnoterule}{}  
\begin{tabular*}{\linewidth}{@{\extracolsep\fill}lrr;c.c..;c.}
\hline
\hline\\
\vspace{-2mm} 
 & 
 & 
 &
\multicolumn{6}{c}{\element[][12]{CO}(1$-$0)} & 
\multicolumn{3}{c}{\element[][12]{CO}(2$-$1)} \\
 & 
 & 
 &
\multicolumn{6}{c}{\downbracefill} & 
\multicolumn{3}{c}{\downbracefill} \\
\multicolumn{1}{c}{Obj.} & 
\multicolumn{1}{c}{$v_0$ (LSR)} &
 \multicolumn{1}{c}{$D_{\rm L}$} &
\multicolumn{1}{c}{$I_{\rm CO} \;$\footnotemark[1]}  & 
\multicolumn{1}{c}{$\Delta v_{\rm FWZI}$} &
\multicolumn{1}{c}{$v_{\rm res}$} &
\multicolumn{1}{c}{$LS \;$\footnotemark[2]} &
\multicolumn{1}{c}{$L^\prime_{\rm CO}$/$10^8 \;$\footnotemark[3] }&
\multicolumn{1}{c}{$M(\rm H_2)$} &
\multicolumn{1}{c}{$I_{\rm CO}\;$\footnotemark[1]}  & 
\multicolumn{1}{c}{$\Delta v_{\rm FWZI}$} &
\multicolumn{1}{c}{$v_{\rm res}$}\\
     &  
 \multicolumn{1}{c}{[km/s]} &
 \multicolumn{1}{c}{[Mpc]} &
 \multicolumn{1}{c}{[K km/s]} & 
 \multicolumn{1}{c}{[km/s]} &
 \multicolumn{1}{c}{[km/s]} &
 &
 \multicolumn{1}{r}{[K km/s pc$^2$]}  &
 \multicolumn{1}{c}{$\rm 10^9 M_\odot$ ]} &
 \multicolumn{1}{c}{[K km/s]} & 
 \multicolumn{1}{c}{[km/s]} &
 \multicolumn{1}{c}{[km/s]}\\
\hline\\
HE 0021-1819      &       15954  & 215.5 & 0;39 \pm 0.04  &   110 & 11.0 & $\wedge$    & 2.1  & 0.8 & 0;68 \pm 0.1     &  121 & 11.0 \\ 
HE 0040-1105      &       12578  & 169.4 & 0;96 \pm 0.08  &   238 & 10.8 & $\sqcap$    & 3.2  & 1.3 & 1;19 \pm 0.15    &  238 & 22.0 \\  
HE 0045-2145      &        6403  &  85.8 & 8;20 \pm 0.18  &   133 &  9.0 & $\wedge$    & 7.2  & 2.9 &22;39 \pm 0.59    &  181 &  5.3 \\  
HE 0108-4743\footnotemark[4]                                                                                        
                  &        7175  &  96.2 & 2;24 \pm 0.07  &   118 &  7.4 & $\sqcap$    &10.4  & 4.2 & 3;67 \pm 0.1     &  160 &  7.4\\
HE 0114-0015      &       13682  & 184.5 & 0;92 \pm 0.13  &   326 & 21.8 & $\sqcap$    & 3.6  & 1.5 & 2;24 \pm 0.38    &  305 & 43.5 \\
HE 0119-0118      &       16412  & 221.7 & 2;38 \pm 0.09  &   165 &  5.5 & $\wedge$    &13.5  & 5.4 & 4;70 \pm 0.17    &  143 &  5.5 \\  
HE 0150-0344      &       14329  & 193.3 & 0;89 \pm 0.15  &   305 & 21.8 & $\sqcap$    & 3.8  & 1.5 &\multicolumn{1}{c}{\footnotemark[7]}&      &      \\  
HE 0212-0059      &        7921  & 106.3 & 3;51 \pm 0.11  &   240 &  5.3 & $\sqcap$    & 4.7 \footnotemark[8] 
                                                                                              & 1.9 \footnotemark[8]
											            & 2;42 \pm 0.12    &  251 &  5.3 \\  
HE 0224-2834      &       18150  & 245.6 & 0;71 \pm 0.11                      
                                        \footnotemark[5]  &   132 & 21.1 & $\wedge$    & 4.9  & 2.0 & <0;46            & 132  & 22.1 \\
HE 0227-0913      &        4914  &  65.8 & 1;86 \pm 0.17  &   254 & 10.6 & $\sqcap$    & 1.0  & 0.4 & 2;96 \pm 0.31    &  254 & 21.1 \\  
HE 0232-0900      &       12886  & 173.6 & 6;92 \pm 0.34  &   597 & 10.9 & $\sqcap$    &24.3  \footnotemark[9]
                                                                                              & 9.7 \footnotemark[9]
											            & 8;13 \pm 0.66    &  586 & 22.0 \\  
HE 0253-1641      &        9580  & 128.7 & 2;89 \pm 0.15  &   282 & 13.4 & $\sqcap$    & 5.6  & 2.3 & 3;89 \pm 0.37    &  301 & 21.0 \\  
HE 0433-1028      &       10651  & 143.3 & 9;42 \pm 0.27  &   252 &  3.4 & $\wedge$    &22.6  & 9.0 &16;55 \pm 0.48    &  199 &  5.4 \\  
HE 0853-0126      &       17899  & 242.1 & 1;28 \pm 0.15  &   331 & 22.1 & $\sqcap$    & 8.6  & 3.4 & 0;82 \pm 0.36    &  309 & 44.1 \\  
HE 0949-0122      &        5905  &  79.1 & 1;23 \pm 0.14  &   276 & 21.2 & $\sqcap$    & 0.9  & 0.4 & 3;55 \pm 0.24    &  318 & 21.2 \\  
HE 1011-0403      &       17572  & 237.6 & 1;37 \pm 0.12  &   187 & 11.0 & $\sqcap$    & 8.9  & 3.5 &\multicolumn{1}{c}{\footnotemark[7]}&      &      \\  
HE 1017-0305      &       14737  & 198.9 & 1;25 \pm 0.10  &   458 & 21.8 & $\sqcap$    & 5.7  & 2.3 & 2;44 \pm 0.23    &  458 & 21.8 \\  
HE 1029-1831\footnotemark[6]                                                                                        
                  &       12112  & 163.1 & 4;24 \pm 0.2   &   254 &  3.4 & $\wedge$    &13.1  & 5.3 & 5;05 \pm 0.2     &  254 &  3.4 \\
HE 1107-0813      &       17481  & 236.4 & 0;82 \pm 0.10  &   308 & 22.0 & $\sqcap$    & 5.2  & 2.1 &<0;3              &  310 & 44.0 \\		  
HE 1108-2813      &        7198  &  96.5 & 8;32 \pm 0.28  &   309 &  5.3 & $\sqcap$    & 9.2  & 3.7 &18;84 \pm 0.47    &  320 &  5.3 \\  
HE 1126-0407      &       18006  & 243.6 & 1;52 \pm 0.15  &   640 & 22.1 & $\sqcap$    & 10.3 & 4.1 & 3;02 \pm 0.20    &  551 & 22.1 \\  
HE 1237-0504      &        2531  &  33.8 & 7;52 \pm 0.57  &   302 & 13.1 & $\sqcap$    & 1.0  & 0.4 & 6;04 \pm 1.01    &  315 & 21.0 \\  
HE 1248-1356      &        4338  &  58.0 & 2;96 \pm 0.19  &   348 & 10.5 & $\sqcap$    & 1.2  & 0.5 & 4;23 \pm 0.39    &  338 & 21.1 \\
HE 1330-1013      &        6744  &  90.4 & 1;82 \pm 0.16  &   266 & 10.6 & $\sqcap$    & 1.8  & 0.7 & 2;4  \pm 0.2     &  245 & 10.6 \\
HE 1353-1917      &       10472  & 140.8 & 3;25 \pm 0.22  &   592 & 10.7 & $\sqcap$    & 7.6  & 3.0 & 5;5  \pm 0.3     &  549 & 10.8 \\
HE 2211-3903\footnotemark[4]                                                                                       
                  &       11945  & 160.8 & 1;87 \pm 0.11  &   466 & 30.0 & $\sqcap$    &23.2  & 9.3 & 1;85 \pm 0.06    & 260  & 15.1 \\
HE 2222-0026      &       17414  & 235.5 & 0;21 \pm 0.04  &   154 & 22.0 & $\wedge$    & 1.3  & 0.5 & 0;61 \pm 0.06    & 176  & 22.0 \\
HE 2233+0124      &       16913  & 228.6 & 0;89 \pm 0.08  &   506 & 22.0 & $\sqcap$    & 5.4  & 2.1 & 0;95 \pm 0.12    & 506  & 22.0 \\
HE 2302-0857      &       14120  & 190.4 & 2;31 \pm 0.16  &   653 & 10.9 & $\sqcap$    & 9.7  & 3.9 & 5;17 \pm 0.29    & 817  & 10.9 \\
\hline
\multicolumn{11}{c}{Non-detections}\\ 
\hline\\
HE 0021-1810      &       16039  &   216.6 & < 0;07           &  315 &  43.8 & &< 1.1 & < 0.5 & <0;12 &      & 43.8\\
HE 0203-0031      &       12723  &   171.4 & < 0;16           &  315 &  43.4 & &< 1.6 & < 0.6 & <0;51 &      & 43.4\\
HE 0345+0056      &        8994  &   120.8 & < 0;15           &  315 &  42.9 & &< 0.8 & < 0.3 & <0;26 &      & 42.9\\
HE 0351+0240      &       10793  &   137.0 & < 0;19           &  315 &  43.1 & &< 1.2 & < 0.5 & \multicolumn{1}{c}{\footnotemark[7]}&      &   \\
HE 0412-0803      &       11392  &   152.9 & < 0;14           &  315 &  43.2 & &< 1.2 & < 0.5 & \multicolumn{1}{c}{\footnotemark[7]}&      &   \\
HE 0429-0247      &       12441  &   165.5 & < 0;31           &  315 &  43.4 & &< 2.9 & < 1.2 & \multicolumn{1}{c}{\footnotemark[7]}&      &   \\
HE 0853+0102      &       15589  &   210.5 & < 0;16           &  315 &  43.8 & &< 2.4 & < 1.0 & \multicolumn{1}{c}{\footnotemark[7]}&      &   \\
HE 0934+0119      &       15091  &   203.7 & < 0;12           &  315 &  43.7 & &< 1.7 & < 0.7 & <0;24 &      & 43.7\\
HE 1310-1051      &       10193  &   137.0 & < 0;16           &  315 &  43.0 & &< 1.0 & < 0.4 & <0;23 &      & 43.0\\
HE 1338-1423      &       12528  &   168.8 & < 0;15           &  315 &  43.4 & &< 1.5 & < 0.6 & <0;21 &      & 43.4\\
HE 1417-0909      &       13191  &   177.8 & < 0;09           &  315 &  43.4 & &< 1.0 & < 0.4 & <0;18 &      & 43.4\\
HE 2128-0221      &       15828  &   213.8 & < 0;08           &  315 &  43.8 & &< 1.3 & < 0.5 & <0;12 &      & 43.8\\
\hline
\end{tabular*}\\
\end{center}
\footnotemark[1] Upper limits and errors represent 1$\sigma$.
\footnotemark[2] Line shape. $\wedge$ denotes a triangular shaped line profile, as it is expected for turbulent line-of-sight velocity dispersion. $\sqcap$ represents boxy or double-horn line profiles indicative for emission from an inclined, rotating disk.    
\footnotemark[3] For the upper $L_{\rm CO}^\prime$ limits, the average linewidth $\Delta v=315 $km s$^{-1}$ and 3$\sigma$ detection limits were considered.
\footnotemark[4] SEST 15m telescope measurement. A beam efficiency $B_{\rm eff}=0.7$ was applied to $T_{\rm A}^*$ and a 45\arcsec\, beam was considered for L$^\prime_{\rm CO}$.
\footnotemark[5] 3$\sigma$ detection level for $T_{\rm mb}$.
\footnotemark[6] IRAM PdBI measurement \citep{2007a&a...464..187k}, resulting in $S\Delta v=(21\pm 1)$Jy km s$^{-1}$ for \element[][12]{CO}(1$-$0) and $S\Delta v=(25\pm 1)$Jy km s$^{-1}$ for \element[][12]{CO}(2$-$1). $S/T_{\rm mb}$ = 4.95 Jy/K was used to compare the PdBI results with the 30m telescope data.
\footnotemark[7] 1mm data not reliable due to high water vapor.
\footnotemark[8] Source is likely to be extended. Value probably too low by an estimated factor 1.7
\footnotemark[9] Source is not compact. Values too low by a factor $\sim$1.7

\end{minipage}
\end{table*}

The CO line luminosity $L^\prime_{\rm CO}$  of the \element[][12]{CO}(1$-$0) transition can be expressed as \citep*{1992apj...398l..29s}:
$$ {L^\prime_{\rm CO}=23.5 \;\Omega_{S\star B}\;D_L^2 \; I_{\rm CO}\; (1+z)^{-3}\;\rm \left[K\; km\; s^{-1}\; pc^2\right] },$$
i.e., as a function of the velocity-integrated line intensity $I_{\rm CO}$ in units of K km s$^{-1}$, the luminosity distance  
${ D_{\rm L}}$ measured in Mpc, and the solid angle of the source convolved with the telescope beam ${\rm \Omega_{S\star B}}$ in square arcseconds.
\
For sources with solid angles much smaller than the telescope beam, ${\rm \Omega_{S\star B}}$ can be approximated with ${\rm \Omega_{Beam}}$. 
\
The expression $L^\prime_{\rm CO}$, commonly used in radio astronomy, and the general expression of the  line luminosity $L_{\rm CO}$ are related by $L_{\rm CO}=\left(8\pi k \nu_{\rm rest}^3/c^3\right)L^\prime_{\rm CO}$ \citep{1992natur.356..318s}. 1K km s$^{-1}$ corresponds to $\sim1.9\cdot10^{29}\textrm{erg s}^{-1}$ at the 115~GHz transition.   

Established correlations \citep{1995ApJS...98..219Y} between the 25.0 B$_{\rm{mag}}$ arcsec$^{-2}$ isophotal diameter $D_{25}$ of galaxies and the size of their CO emission region were used to confirm that, with the exception of \object{HE~0212-0059} and \object{HE~0232-0900}, the sources should all be smaller than the $\sim$22\arcsec telescope beam resulting at $\sim$115 GHz. For the two exceptional cases it is likely that the overall CO line luminosity is underestimated in the present study. In the case of \object{HE~0232-0900}, already published data \citep[][see below]{1995a&a...298..743h} allowed to determine a factor $\sim$1.7 by which the current data is too low. Because of the very similar optical extent, a similar  factor can be assumed for  \object{HE~0212-0059}.

The integrated line intensity, the systemic velocity (represented by the flux-weighted centroid of the CO(1-0) line),  the linewidth (FWZI), the resolution of the smoothed spectrum as well as the derived CO luminosity of each detected source is shown in Table \ref{TabResults}.  
\
Also shown are estimates for the molecular gas masses. They were obtained by applying a $M({\rm H_2}) / L_{\rm CO}^\prime$ conversion factor $\alpha$ to the CO luminosity. 
\
The determination of the H$_2$ content using $L^\prime_{\rm CO}$ as tracer has to be handled with care. 
\
Commonly used is a value of $\alpha$=4.8 $\rm M_\odot$ (K km s$^{-1}$ pc$^2$)$^{-1}$ \citep{1991iaus..146..235s}, derived from galactic molecular cloud observations.
\
The common approach of using a 'standard' conversion factor derived from galactic observations has the shortcoming of disregarding the dependency of $\alpha$ on the metalicity of the region of interest \citep{1997A&A...328..471I}.
\
For a sample of ultra luminous infrared galaxies \citet{1998apj...507..615d} state a conversion factor that is 5 times lower. 
\.
Within this paper $\alpha=4$ $\rm M_\odot$ (K km s$^{-1}$ pc$^2$)$^{-1}$ is adopted to allow for direct comparison with the results of \citet{2001aj....121.3285e, 2006aj....132.2398e} or \citet{2003apj...585l.105s}.

A subset of the sources had been subject to molecular gas studies before. The published data confirms the results listed in Table \ref{TabResults}.
\
The \element[][12]{CO}(1$-$0) emission of the extended source \object{HE~0232-0900} was mapped with the IRAM 30m telescope by \citet{1995a&a...298..743h}. The value for $I_{\rm CO}$ of the central position agrees within 18\%. 
\
After correction for the different beam sizes, the \element[][12]{CO}(1$-$0) detections of \object{HE~0227-0913} with the NRO 45m telescope by \citet{1998aj....116.1553v},  of \object{HE~1237-0504} with SEST by \citet{2000a&as..141..193c}, and of \object{HE~0433-1028} with SEST by \citet{2004mnras.353.1151s} agree within 10\% with the values presented in this paper. 
\
\object{HE~1237-0504} was also observed by \citet{1997apj...485..552m} with the NRAO 12m telescope. Their beam-size corrected values for this object, for \object{HE~0212-0059}, and for \object{HE~0949-0122}, however, exceed the values presented here by a factor 2-3. For \object{HE~0212-0059}, the difference may be explained with the extension of the galaxy and the different beam sizes. The other results, however, are in contrast with the good agreements of the 4 other independent measurements mentioned above, which provide a high level of confidence in the data presented here.

\section{Discussion}
\label{SectDiscussion}
\subsection{Detection rate and molecular gas mass}
\label{SectDetect}
\begin{figure}
  \resizebox{\hsize}{!}{\includegraphics{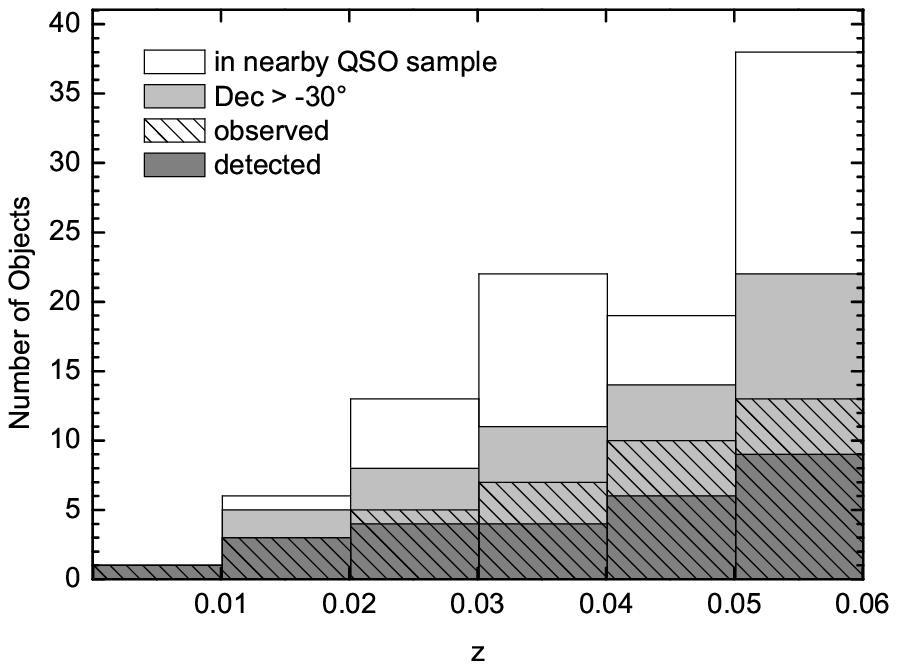}}
  \caption{The number of existing / observable with the IRAM 30m telescope / observed / detected HE~QSOs with increasing volume. At the distances discussed here, the frequency of occurrence is dominated by the size of the comoving volume.}
  \label{FigZDistrib}
\end{figure}
About 60\% of the mostly southern sources in the nearby low-luminosity QSO sample have a declination $\ge -30^\circ$ and can be observed with the IRAM 30m telescope. Of these, about 65\% have been subject to our study. They show a redshift distribution similar to the full nearby low-luminosity QSO sample (Fig.  \ref{FigZDistrib}). Within the restricted redshift range considered in this sample, the number of objects per redshift bin is purely volume dependent -- the number increases with increasing z.
\begin{figure}
  \resizebox{\hsize}{!}{\includegraphics{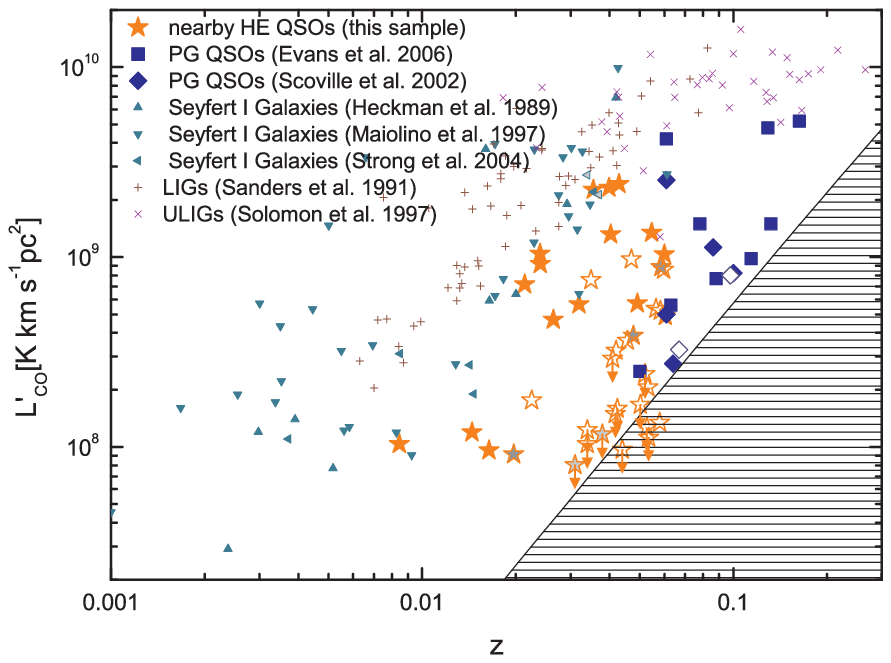}}
  \caption{CO luminosity versus redshift. For objects represented by filled shapes $L_{\rm FIR}$ is available, grey shapes indicate available upper limits for $L_{\rm FIR}$ and outlined shapes represent objects that have not been listed in the IRAS FSC. The shaded area in the bottom right corner indicates the region in which the determination of $I_{\rm CO}$ with the 30m telescope  falls below 0.315 K km s$^{-1}$. This value corresponds to a signal of 1 mK with a linewidth of 315 km s$^{-1}$, the average width of the detected lines. \emph{[See the online edition of the Journal for a color version of this figure.]}} 
  \label{FigLCOz}
\end{figure} 
\
When plotted against redshift (Fig. \ref{FigLCOz}), the particular position of the sample in between previously studied local Seyfert I and more luminous QSO host galaxies becomes apparent.
\
The omission of a far infrared (FIR) selection and the on average larger distance compared to the local Seyfert I population resulted in the inclusion of several objects  with FIR flux densities below the IRAS detection limit (cf. Sect. \ref{SectFIR}).     
\
In several of these cases we still were able to detect CO emission.  

Our 3$\sigma$ detection limits agree with the limits in the PG QSO host studies by \citet{2001aj....121.3285e,2006aj....132.2398e} and  \citet{2003apj...585l.105s}, carried out with the OVRO array and the IRAM 30m telescope. 
\
Different from \citet{2001aj....121.3285e,2006aj....132.2398e}, the volume limited sample of \citet{2003apj...585l.105s}, consisting of 12 objects, is not confined to FIR selected sources. Therefore, a direct comparison with their results is appropriate.
\
At redshifts exceeding 0.06, the common detection limit prevents the detection of galaxies with a molecular gas content of below $10^9$~M$_{\odot}$ -- only gas rich objects are then detectable. The redshifts of virtually all PG QSOs exceed this value. 
\
Nevertheless, sample of \citet{2003apj...585l.105s} resulted in a detection rate of 75\% which lets the authors conclude that the majority of   
luminous, low-redshift QSOs have gas-rich host galaxies. 
\
This picture may become more complicated when considering the incompleteness of the PG survey at low redshifts. The rejection of extended objects may cause an underrepresentation of bulge-dominated galaxies with a lower gas content. 
\
On the other hand, due to the arbitrariness of the $M_{\rm B}\sim-22$ borderline, low-luminosity QSOs above the line still may have host properties that are similar to their brighter equivalent. 
\
With 70\% the detection rate of the IRAM 30m observations presented here is almost identical to the rate of \citet{2003apj...585l.105s}.
\
The molecular gas masses of the detected host galaxies range from 0.4$\cdot$10$^9\rm M_{\odot}$ to 9.7$\cdot$10$^9\rm M_{\odot}$ with an average molecular gas mass of 2.8$\cdot$10$^9\rm M_{\odot}$ (3.0$\cdot$10$^9\rm M_{\odot}$ when including the 2 sources that were detected with SEST).
\
\subsection{Comparison of CO and optical redshifts}
\label{SectRedshift}
\begin{figure}
  \resizebox{\hsize}{!}{\includegraphics{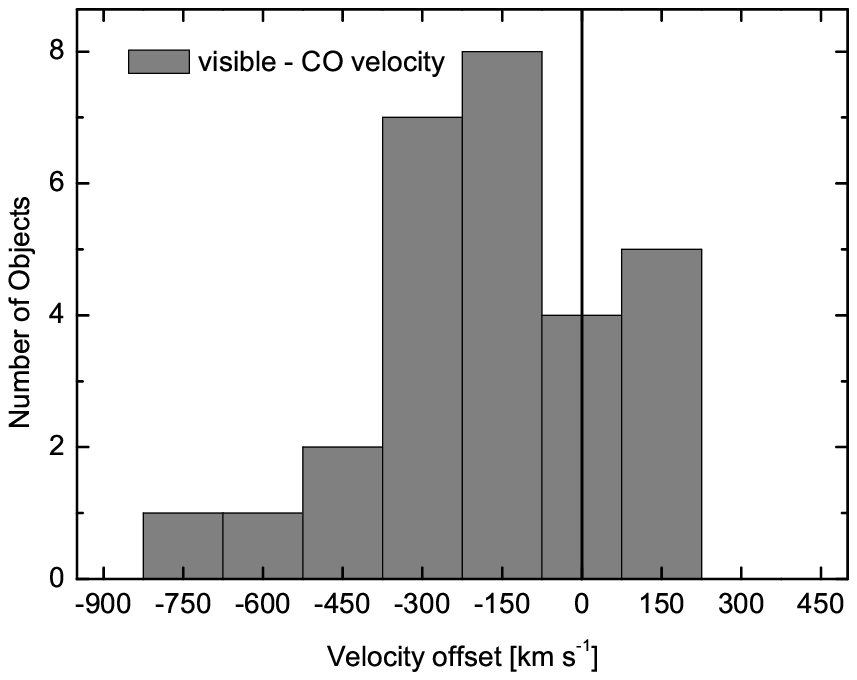}}
  \caption{The distribution of the differences between the CO velocity and the velocity derived from restframe visible spectroscopy. The systematic blueshift of the spectroscopic features determined in the visible is an indication for outflows or partial dust obscuration in the vicinity of the active nucleus.} 
  \label{FigvelDrift}
\end{figure} 
\
Millimetric CO observations are well suited for the accurate measurement of systemic velocities of active galaxies. CO traces the cold and extended molecular gas distribution of the total host galaxy rather than the reservoir of highly excited gas in proximity to the central source. Features of the latter are commonly used to determine the redshift of active galaxies in the visible wavelength domain. These redshifts, however, can be affected by outflows, dust obscuration or other asymmetric phenomena that result in apparent velocities deviating from the actual systemic velocity. 

For nearby objects like the sample discussed in this paper, the redshifts determined in the course of the HES followup spectroscopy program are mainly based on the narrow [\ion{O}{iii}]${\lambda5007}$ emission line \citep{1996a&as..115..235r}. \citet{2005aj....130..381b} shows that the [\ion{O}{iii}] line is blue-shifted relative to the systemic velocity  in up to 50\% of all AGN. The emission peak shift may be as large as several hundred km s$^{-1}$. Fig. \ref{FigvelDrift} shows the distribution of offsets between the velocities corresponding to the HE redshifts and the CO velocities measured in this study. Negative velocity differences imply blue-shifted visible emission features. On top of the nominal uncertainty of HE velocities ($\sim$300 km s$^{-1}$), the nearby low-luminosity QSOs indeed show a tendency for such a blue-shift with a mean velocity offset of $-174$ km s$^{-1}$.

\subsection{Blue luminosity and CO line characteristics}
\begin{figure}
  \resizebox{\hsize}{!}{\includegraphics{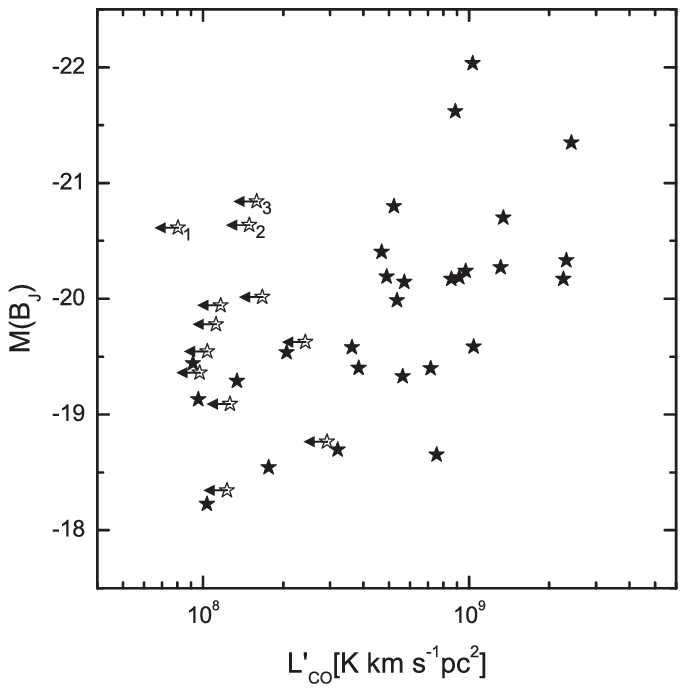}}
  \caption{Absolute $B_J$ magnitude versus CO luminosity. The magnitudes shown here represent the central seeing disk, including contributions of both AGN and star formation in the central part of each galaxy. The majority of objects showing $M(B_J)$ brighter than -20 mag are also CO luminous. Exceptions are (1) \object{HE~0345+0056}, (2) \object{HE~1338-1423}, and (3) \object{HE~0203-0031}.}   
  \label{FigLCOMB}
\end{figure}
\begin{figure}
  \resizebox{\hsize}{!}{\includegraphics{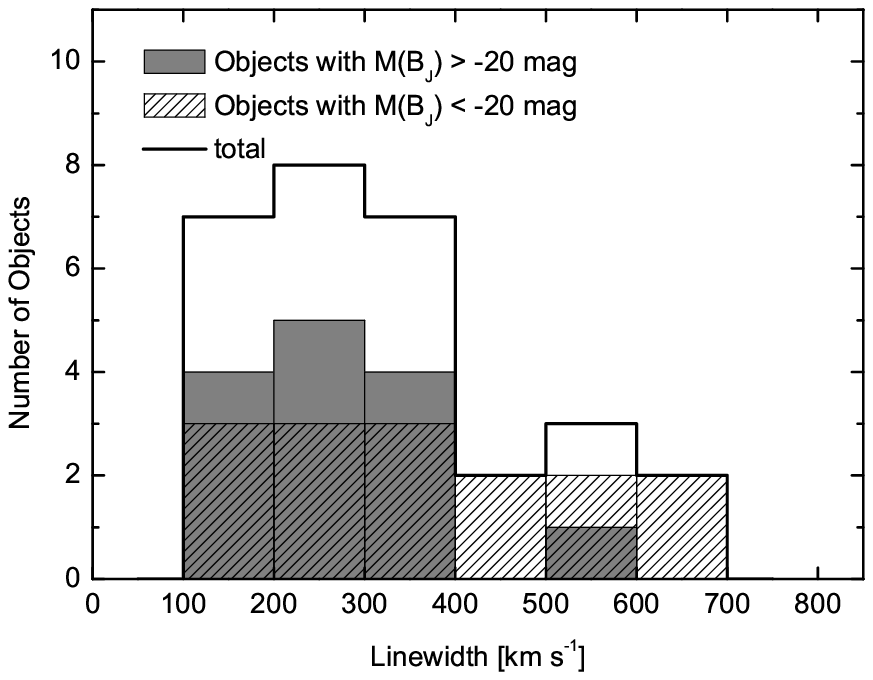}}
  \caption{Distribution of CO line widths (FWZI). Linewidths exceeding $400$ km s$^{-1}$ can be commonly found only in bright objects with absolute blue magnitudes $<-20$.}  
  \label{FigNCLW}
\end{figure}
\
\begin{table}
\begin{center}
\caption{Mean CO(1--0) line widths and standard deviations of the mean for different line shapes and different absolute blue magnitudes. The numbers in parentheses represent the number of objects that fall into the corresponding category.}
\label{TabLW}
\begin{tabular*}{\columnwidth}{@{\extracolsep\fill}lrrr}
\hline
\hline
\multicolumn{1}{c}{} &
\multicolumn{1}{c}{$\wedge$} &
\multicolumn{1}{c}{$\sqcap$} &
\multicolumn{1}{c}{$\wedge + \sqcap$} \\
\hline
 $M(B_J)>-20$ & $132\pm 13$ (3)  &  $301 \pm 34$ (11) &  $265\pm 33$ (14)   \\ 
 $M(B_J)<-20$ & $200\pm 28$ (4)  &  $426 \pm 49$ (11) &  $367\pm 45$ (15)   \\
 all          & $171\pm 22$ (7)  &  $364 \pm 32$ (22) &  $317\pm 29$ (29)   \\
\hline
\end{tabular*}\\
\end{center} 
\end{table}

Fig. \ref{FigLCOMB} shows the connection between  absolute $B_J$  brightness and CO luminosity. The $B_J$ fluxes have a photometric uncertainty of $\sim$0.2~mag and were determined by the HES group, based on their spectral plate data. Each value represents the flux within the central seeing disk, which for type I AGN at these redshifts is dominated by the nuclear contribution, with only small proportions from the host galaxy.

A more detailed description is given in \citet{2000a&a...358...77w}. The magnitudes shown here are not corrected for extinction and, therefore, systematically too faint by a mean value of $\sim$0.23~mag. 

The data does not allow to distinguish between host and AGN contribution. Thus an increase of $B_J$ luminosity can be the result of enhanced star formation activity in the central 1-2 kpc, enhanced AGN activity or both. In any way, the increase in activity comes along with larger reservoirs of molecular gas, as can be seen in Fig. \ref{FigLCOMB}. Except for 3 non-detected sources, low-luminosity QSOs with absolute central  magnitudes brighter than $M(B_J)=-20$~mag all have CO luminosities of $\ge 5\cdot10^8$ K km s$^{-1}$ pc$^2$, while the majority of fainter objects contain less molecular gas. This correlation seems to be intrinsic and not the result of a distance selection bias (with more luminous AGN being at larger distances with higher CO detection limits), as the small fraction of non-detections in the bright case shows. It remains unclear, whether the 3 $B_J$ luminous non-detections are actually gas depleted  or whether they were not detected because of an uncertain CO redshift. Upcoming HI observations, but also analyses of the  morphologies in the visible/NIR will provide further information on the nature of these objects.  

Not only the total CO content is higher but also the spread of the CO(1$-$0) line widths is wider for the brighter AGN hosts. Fig. \ref{FigNCLW} shows the distribution of line widths in all detected galaxies. With one exception, all objects showing FWZI line widths broader than 400 km s$^{-1}$ have absolute central magnitudes of $M(B_J)\le-20$~mag. The average line width of these bright objects is  367~km s$^{-1}$, $\sim$100~km~s$^{-1}$ broader than the fainter sources (cf. Tab. \ref{TabLW}). 
\
The average signal-to-noise ratio (S/N) of the CO(1$-$0) detections in the $B_J$ luminous case is only slightly (1.2 times) larger than the average S/N in the fainter case. Therefore, a significant S/N bias that potentially favors the identification of broader line profiles in more active objects is not present.

Such a different behavior could result from a selection effect in the optical identification process of QSO candidates. Objects with faint AGN activity may only be identified as type I AGN host when seen almost face-on, leaving the Seyfert I nucleus unobscured, while brighter AGN contributions can also be identified in objects with higher inclination.
\
A scenario of an intrinsically broader line width distribution, however, cannot be ruled out. A separation of sources into a class of triangular shaped and a second class of box shaped or double horn line profiles resulted in similar counts for each class (cf. Tables \ref{TabResults} and \ref{TabLW}) in both the faint and the bright cases. The triangular shaped lines are indicative for a face-on view onto a turbulent gas distribution without a dominating disk component in the  line-of-sight direction. The second class represents objects with rotating CO emission disks or rings at a certain inclination with respect to the line-of-sight. In several low-signal cases the assignment of a class was not obvious. Nevertheless,  an aforementioned selection effect that favors low-inclination views on faint AGN candidates should yield a higher fraction of triangular shaped line profiles, which is not the case. Noting the caveat of small number statistics, we would like to point out the difference in the mean line widths of triangular shaped line profiles. Bright objects with triangular shaped line profiles have a mean line width of 200 km~s$^{-1}$, whereas in the faint cases the mean line width is only $\sim$130 km~s$^{-1}$. This difference could be a sign of higher velocity dispersion in the molecular gas distribution of the more active objects.       

Most of the line profiles do not show strong asymmetries that can be used as indication for ongoing interaction. Clear asymmetries can be seen in the two sources that are probably larger than the telescope beam (\object{HE~0212-0059} and \object{HE~0232-0900}). In these cases pointing errors are the likely explanation for the asymmetric line profile.

\subsection{Far infrared luminosity and star formation efficiency}
\label{SectSFE}
\begin{table}
\begin{minipage}[t]{\columnwidth}
\begin{center}
\renewcommand{\footnoterule}{}  
\caption{Far infrared properties of the IRAS detected sources in the subsample. The 60 and 100 $\mu m$ IRAS flux densities are taken from the IRAS Faint Source Catalog \citep{1990irasf.c......0m}, or in few cases from \citet{1988apjs...68...91r}, \citet{1989apj...347...29s}, or \citet{2003aj....126.1607s}. 
}
\label{TabFIR}
\begin{tabular*}{\columnwidth}{@{\extracolsep\fill}lrrr}
\hline
\hline
\multicolumn{1}{c}{Obj.} &
\multicolumn{1}{c}{$F_{\rm 60\mu m}$} &
\multicolumn{1}{c}{$F_{\rm 100\mu m}$} &
\multicolumn{1}{c}{$L_{\rm FIR}$} \\
&
\multicolumn{1}{c}{[Jy]} &
\multicolumn{1}{c}{[Jy]} &
\multicolumn{1}{c}{[$10^{10} L_{\odot}$]} \\
\hline
 HE 0045-2145 &  3.6    &  5.3    &    4.2 \\ 
 HE 0108-4743 &  1.0    &  2.2    &    1.7 \\         
 HE 0119-0118 &  1.5    &  1.8    &   10.8 \\
 HE 0150-0344 &  0.5    & $<$1.7  & $<$4.4 \\  
 HE 0212-0059 &  0.5    &  1.5    &    1.2 \\
 HE 0224-2834 &  0.4    &  0.4    &    3.3 \\ 
 HE 0227-0913 &  0.4    &  0.9    &    0.3 \\ 
 HE 0232-0900 &  1.4    &  1.9    &    6.5 \\ 
 HE 0253-1641 &  0.7    &  0.8    &    1.6 \\ 
 HE 0345+0056 &  0.5    & $<$3.2  & $<$2.7 \\ 
 HE 0412-0803 &  0.6    & $<$1.4  & $<$2.8 \\ 
 HE 0433-1028 &  2.7    &  4.2    &    9.1 \\  
 HE 0949-0122 &  1.3    & $<$2.4  & $<$1.5 \\
 HE 1011-0403 &  0.2    & $<$0.3  & $<$1.5 \\
 HE 1017-0305 &  0.5    &  0.6    &    2.8 \\ 
 HE 1029-1831 &  2.6    &  3.7    &   10.8 \\
 HE 1108-2813 &  3.1    &  4.2    &    4.5 \\
 HE 1126-0407 &  0.7    &  1.2    &    6.8 \\
 HE 1237-0504 &  3.1    &  6.0    &    0.6 \\
 HE 1248-1356 &  0.8    &  1.3    &    0.5 \\       
 HE 2211-3903 &  0.8    &  1.1    &    3.2 \\

\hline
\end{tabular*}\\
\end{center} 
\end{minipage}
\end{table}

The correlation between molecular gas and far infrared emission is well known and often discussed in the context of CO studies.  $L_{\rm FIR}/L'_{\rm CO}$ or $L_{\rm FIR}/M({\rm H_2})$ is commonly referred to as Star Formation Efficiency (SFE) indicator, assuming that $L'_{\rm CO}$ traces the cold, gravitationally bound molecular gas reservoirs that form the birth places of young, dust enshrouded stars. However, CO is only an indirect measure as it actually traces the overall molecular gas content in galaxies rather than explicitly the dense cores that produce young stars \citep[e.g.,][]{2004apj...606..271g}. Furthermore, the far IR emission of galaxies is a composite of an active star formation and a quiescent cirrus-like component \citep{1986apj...311l..33h}. In the case of AGN hosts, AGN driven dust heating may also contribute significantly to the FIR luminosity. 

Many samples are based on IR selection criteria and, therefore, represent only the IR and CO bright tail of the luminosity distribution.
\
So do several studies of the molecular gas content of Seyfert galaxies  \citep[e.g.][]{1989apj...342..735h,1998apj...492..521p,2004mnras.353.1151s} or the study of PG QSO host galaxies by \citet{2001aj....121.3285e,2006aj....132.2398e}. 
\
For the sample discussed in this paper an IR selection criterion was not introduced.  
\
Only $\sim$50\% of the nearby HE QSOs sample are listed in IRAS catalogs. The subsample discussed here contains the same fraction of IRAS sources (cf. Table \ref{TabFIR}) -- an IR bias was carefully avoided. 
\
65\% of the detected sources are listed in IRAS catalogs, while only 2 of the 12 non-detections are detected by IRAS. Only \object{HE~2302-0857} is located in a field that was not scanned by IRAS.     

Many of the IRAS detected sources only show 60 $\mu$m and 100 $\mu$m flux densities and several of them are close to the IRAS detection limit. Due to the limited data at hand, the following discussion will focus on $L_{\rm FIR}$. It is based on the 60 $\mu$m and 100 $\mu$m IRAS flux densities and represents the luminosity between 43$\mu$m and 123$\mu$m. Due to this bandwidth limit, warmer components peaking at shorter wavelengths, especially  potential AGN heated dust components remain less considered in $L_{\rm FIR}$. In the context of LIGs/ULIGs or nearby galaxies, $L_{\rm IR}$ is often used instead of $L_{\rm FIR}$. $L_{\rm IR}$ also includes the 12$\mu$m and 25$\mu$m IRAS bands and represents the total IR luminosity ranging from 8 to 1000 $\mu$m. For the definition of $L_{\rm FIR}$ and $L_{\rm IR}$ cf. \citet{1996ara&a..34..749s}.  
\label{SectFIR}
\begin{figure}
  \resizebox{\hsize}{!}{\includegraphics{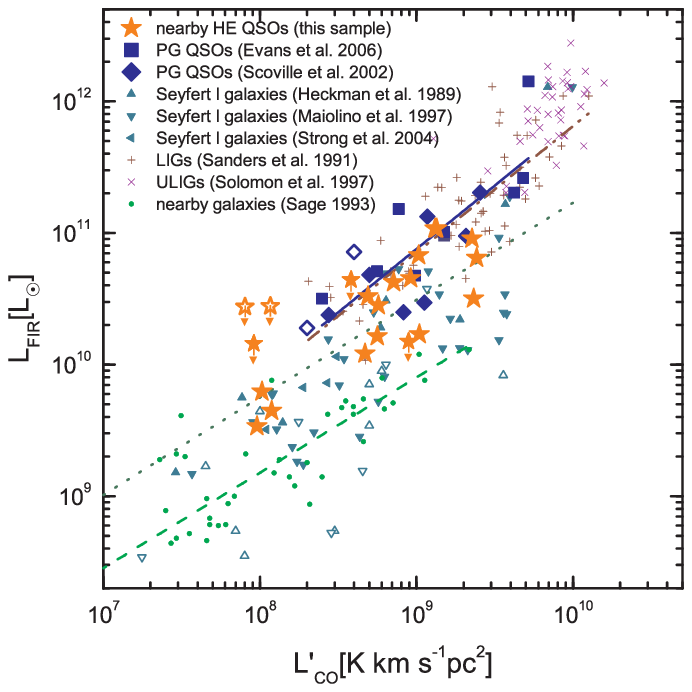}}
  \caption{FIR luminosity as a function of CO luminosity for the nearby QSO sample and various other Seyfert I, PG QSO host, luminous infrared galaxies and normal galaxies. The solid line represents a power law fit to the PG QSO hosts, the dashed line the distance limited sample of normal galaxies of \citet{1993a&a...272..123s}, the dotted line a flux-limited sample of normal spiral galaxies \citep{1988apj...334..613s}, and the dashed-dotted line represents a power-law fit to the LIG sample of \citet{1991apj...370..158s}. Outlined symbols represent upper $L'_{\rm CO}$ limits. \emph{[See the online edition of the Journal for a color version of this figure.]}  }
  \label{FigLCOFIR}
\end{figure}
\\

Fig. \ref{FigLCOFIR} shows the distribution of the IRAS detected sources in the $L'_{\rm CO}-L_{\rm FIR}$ diagram, together with the CO data of a volume limited sample of nearby spiral galaxies \citep{1993a&a...272..123s}, 
Seyfert I galaxies \citep{1989apj...342..735h,1997apj...485..552m,2004mnras.353.1151s},
luminous \citep[LIGs,][]{1991apj...370..158s} and ultraluminous infrared galaxies  \citep[ULIGs,][]{1997apj...478..144s}, and PG QSOs \citep{2003apj...585l.105s,2006aj....132.2398e}.
Also shown are linear regression fits (in log-log space) to normal and to luminous IR galaxy samples as well as to the PG QSO hosts.
\

The distribution of type I AGN hosts in the plot indicates the presence of two populations with differing power-laws:
one population that follows the power-law of normal spiral galaxies and a second population that has $L_{\rm FIR}/L'_{\rm CO}$ properties very similar to LIGs. 

The studies of \citet{1988apj...334..613s} and \citet{1993a&a...272..123s} are used to quantify the power law for normal galaxies:  
A fit to the distance limited sample of normal galaxies (excluding the upper limits considered in \citet{1993a&a...272..123s}) resulted in 
$$L_{\rm FIR}/{10^9} = 8\cdot [L'_{\rm CO}/{10^9}]^{0.72(\pm0.05)},$$
while a fit to a FIR-selected sample of isolated spiral galaxies or spiral galaxies that do not show signs of interaction \citep{1988apj...334..613s} yielded 
$$L_{\rm FIR}/{10^9} = 31\cdot [L'_{\rm CO}/{10^9}]^{0.74(\pm0.07)}.$$ 
Due to the FIR selection criterion, the latter sample is biased towards more active star formation with respect to average spiral galaxies. 
\
The bulk of local Seyfert I galaxies follows the trend described by these two fits, with a somewhat higher $L_{\rm FIR}/L'_{\rm CO}$ ratio for any given  $L'_{\rm CO}$ than the \citet{1993a&a...272..123s} sample.

The power law fit to the PG QSO hosts, on the other hand, produces a slope of about unity:
$$L_{\rm FIR}/{10^9} = 75\cdot [L'_{\rm CO}/{10^9}]^{0.96(\pm0.19)}$$
and is almost identical to the fit to the LIG data:
$$L_{\rm FIR}/{10^9} = 71\cdot [L'_{\rm CO}/{10^9}]^{0.96(\pm0.09)}.$$
\
This common trend may be indicative for similar star formation activity in LIGs and luminous QSO hosts. 

As for the low-luminosity QSO sample, Figure \ref{FigLCOFIR} supports the picture of the sample being a link between the local Seyfert~I population and luminous QSO host galaxies. The sample members seem to follow both trends, most of them can be found in a transition region between the two populations.
\
\subsection{Opaque CO and FIR emission in (U)LIGs and QSO host galaxies?}
\
In the following we will summarize the various published results on the  $L'_{\rm CO}$--$L_{\rm FIR}$ relation in normal and infrared luminous galaxies together with the findings on the nature of the emission regions. We combine  different pieces of information to a simple qualitative model that  explains geometrically the separation of type I AGN hosts into two populations in the $L_{\rm FIR}$ range between $10^9 L_{\odot}$ and $5\cdot10^{11} L_{\odot}$. 

As discussed in Sect. \ref{SectSFE}, normal, undisturbed galaxies show a nonlinear $L'_{\rm CO}$ -- $L_{\rm FIR}$ relation with a slope $<$1. For interacting galaxies \citep{1988apj...334..613s} and LIGs \citep{1991apj...370..158s} the slope is approaching unity. In this case a linear relation between CO and FIR emission can be assumed. For ULIGs \citet{1997apj...478..144s} also find a linear relation between FIR and CO emission, with a higher $L_{\rm FIR}/L_{\rm CO}$ ratio than in normal galaxies and LIGs. The increase of the $L_{\rm FIR}/L_{\rm CO}$ ratio towards ULIGs is well known \citep[e.g.,][]{1986apj...305l..45s, 1996ara&a..34..749s,2004apj...606..271g}. It extends towards the even more luminous SMGs \citep{2005mnras.359.1165g}.    
\citet{2004apj...606..271g} point out, that a linear relation between CO and FIR luminosity remains valid only over two orders of magnitude in the lower $L_{\rm FIR}$ range (while a corresponding HCN -- FIR relation extends over three orders of magnitude and includes the ULIG domain). Any fit to the CO luminosity over a wider range naturally yields a slope $>$ 1. \citeauthor{2004apj...606..271g} state a slope of 1.25$\pm$0.08 for $L_{\rm FIR}$ between 10$^{9.5} L_{\odot}$ and 10$^{12.5} L_{\odot}$. \citet{2003apj...588..771y} even obtain a slope of  1.7 for $L_{\rm FIR}$  between 10$^{9} L_{\odot}$ and 10$^{12} L_{\odot}$. \footnote{Their small projected beam size, however, may not cover the full CO content especially in the nearby, less infrared luminous galaxies. Thus, the slope may appear steeper than it actually is.}               

In interferometric CO data on nearby ULIGs \citet{1998apj...507..615d} show that most of the emission originates from rotating circumnuclear disks or rings. The disk gas forms a continuous medium rather than discrete virialized clouds. 
The linearity between FIR and CO emission in ULIGs and the dense environment resulting from the massive concentration of molecular gas suggest thermal excitation and optically thick emission. This is proposed by the same authors in an earlier paper \citep{1997apj...478..144s}  but partially revised in \citet{1998apj...507..615d}: According to their model fits the CO line emission is only moderately opaque and the CO(2--1)/CO(1--0) ratio indicates sub-thermal excitation. \\

If a connection between ULIGs and QSOs exists, as suggested in the evolutionary model, what does this imply for the molecular gas content in the evolved QSO phase of these objects? The model assumes a dispersion of the majority of gas and interstellar dust to clear the view onto the formerly dust enshrouded AGN. Indeed, low-z PG QSOs and nearby HE QSOs show CO and FIR luminosities and star formation efficiencies that are lower than it is the case for ULIGs in the same comoving volume. Nevertheless, the majority of nearby low-luminosity QSOs is associated with large reservoirs of molecular gas (Sect. \ref{SectDetect}) and presumably also with ongoing star formation, as it is indicated by their  $L'_{\rm CO}$ -- $L_{\rm FIR}$ distribution (Fig. \ref{FigLCOFIR}). This might imply that the density of the formerly AGN obscuring medium in the central region is significantly reduced.

But how does the evolution affect the distribution of molecular gas?
A confinement of the molecular gas to a compact region is still indicated also in high redshift QSOs \citep[][in ground-state transition observations at z$\ga$4]{2006apj...650..604r}. Spatially resolved data on QSOs in the local universe, however, is hardly available. For \object{Mrk~231}  \citet{1998apj...507..615d} report on an inner and an outer face-on disk within a radius of 1.15 kpc. \citet{2004apj...609...85s} present a ringlike structure with a radius of $\sim$ 1.2 kpc in the circumnuclear molecular gas distribution of \object{I~Zw~1}. For the low-luminosity QSO \object{HE~1029-1831} \citet{2007a&a...464..187k} estimates a size of 6$\pm$2 kpc for the CO source that is aligned with the optical bar. The interferometric maps of the low-z PG QSOs by \cite{2001aj....121.3285e} and \cite{2003apj...585l.105s} do not permit to draw conclusions on the compactness of the emission.        

The coincidence with the LIG population in the $\log L'_{\rm CO}$--$\log L_{\rm FIR}$ diagram with a power-law index 1 can be taken as indicator for a confinement of remaining H$_2$ content to a small and dense region. Based on this  assumption, we would like to revisit the idea of both bands being optically thick.  If this is the case, the two different inclinations in the $\log L'_{\rm CO}$--$\log L_{\rm FIR}$ diagram may have a simple geometrical explanation:
For normal galaxies, the CO distribution is less confined. The totality of individual gas clouds with varying line-of-sight velocities contributes to the spectrum. In this case, it is appropriate  to use the velocity-integrated CO line emission as tracer for the total molecular gas content of the galaxy. 
\
The FIR luminosity, on the other hand, is dominated by reradiated dust emission, which originates from the densest regions in the ISM, the seeds of star formation. If the column densities in these compact regions result in $\tau > 1$ at the FIR bands, only the radiating surface contributes to the detectable FIR flux density. 
\
Let us assume a coarse, linear relation between the molecular gas content and the number of star formation seeds in a 3-dimensional gas distribution. Any  surface of optically thick emission then scales with power of $\frac{2}{3}$ of the overall volume. 
\
If $L'_{\rm CO}$ can be used as measure for the total molecular gas content, $L_{\rm FIR}$ should scale with \smash{$L^{\prime0.67}_{\rm CO}$} and produce the corresponding inclination in the log--log plot. 
\
Once the density of the ISM reaches a threshold at which the CO emitting region (which may consist of a continuous medium) becomes opaque, both the CO and the FIR emitting surface scale linearly with the overall molecular gas and dust content. Under these circumstances the slope in the log--log plot is close to unity.  
\
An even higher density of star formation seeds in the ISM results in a larger surface filling factor within the galaxy, allowing for a higher $L_{\rm FIR}/L'_{\rm CO}$ ratio for each given CO luminosity. 
This effect causes the dispersion perpendicular to the power-law fit. As shown in the Appendix the different power-law fits for the individual source classes can be parameterized via the fractional source filling factor $a$ of the material that participates in intense star formation or the source type $f$ (QSO hosts and LIGs/ULIGs or normal galaxies).

The LIG/QSO scenario and the normal galaxy scenario describe the two extreme cases with power-laws 1 and 0.67 as upper and lower limit. These extreme cases seem to correspond well to the two populations described above. Furthermore, these scenarios do not exclude a transition region, in which the dense central region of a galaxy is optically thick for both bands while for the outer regions the full volume contributes to the detected CO emission.                

\section{Summary}
This paper reports on \element[][12]{CO} observations of 41 nearby low-luminosity QSOs. Thirty-nine of them form a subsample that is not biased towards FIR emission. The results of this study can be summarized as follows:
\begin{enumerate}
\item 70\% of the subsample have been detected in the \element[][12]{CO}(1$-$0) transition. With a mean H$_2$ mass  of $2.8\cdot10^9$M$_{\odot}$ (assuming $\alpha=4$ $\rm M_\odot$ (K km s$^{-1}$ pc$^2$)$^{-1}$) the majority of low-luminosity QSO hosts is rich in molecular gas. This confirms previous results on PG QSOs by \citet{2003apj...585l.105s}.
\item For the majority of detected objects, the redshift based on visible features is blueshifted with respect to the CO line centroid -- probably as a result of asymmetric phenomena in the vicinity of the nucleus, the emission region of the visible features.
\item The absolute $B_J$ magnitude of the central seeing disk of the low-luminosity QSOs is used as an indicator for the total activity of AGN and circumnuclear starburst. Objects with $M(B_J)<-20$mag not only show higher CO luminosities but also a wider spread in the distribution of linewidths than the fainter objects.
\item The connection between CO and FIR properties is discussed in the context of normal galaxies, local Seyfert I objects, LIGs, ULIGs and PG QSOs. Two populations with different power-laws in the $L'_{\rm CO}$--$L_{\rm FIR}$ plot are identified in the range between $L_{\rm FIR}=10^9 \rm \bf L_{\odot}$ and  $L_{\rm FIR}=5\cdot10^{11}\rm \bf L_{\odot}$. The first population shows a power-law index of $\sim0.7$ and contains normal, non- or weakly interacting galaxies and the bulk of Seyfert I objects. The second population with a power-law index of $\sim1$ contains LIGs and the PG QSOs. The low-luminosity QSO sample seems to consist of objects of both populations and objects from a transition region. 
\item For Objects of the second population, i.e. also for galaxies hosting brighter QSOs, the linear relation between FIR and CO luminosity can be taken as indication for a concentration of the molecular gas in compact regions, similar to the case of ULIGs. 
\item The idea of thermalized, optically thick FIR and CO emission, previously suggested in the context of ULIGs, is revisited. The two different slopes in the $L'_{\rm CO}$--$L_{\rm FIR}$ plot are discussed as a potential result of different optical properties of the ISM in the two populations.  
\end{enumerate}

\begin{acknowledgements}
      TB would like to thank the staff of the IRAM 30m telescope for their kind support. We thank the anonymous referee for helpful comments. Part of this work was supported by the Deut\-sche For\-schungs\-ge\-mein\-schaft (DFG), project number: SFB 494. 
\end{acknowledgements}

\begin{appendix}

\noindent
\FloatBarrier
\section{Descriptive star formation model}
\begin{figure}
  \resizebox{\hsize}{!}{\includegraphics{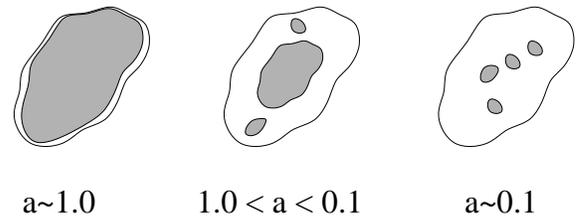}}
  \caption{Examples of molecular gas distributions with different fractions of molecular gas taking part in active star formation (dark areas). We show an inclined host and the outer continuous line includes the majority of the molecular gas in the system.}
  \label{FigActivity}
\end{figure}
\
\begin{figure}
  \resizebox{\hsize}{!}{\includegraphics{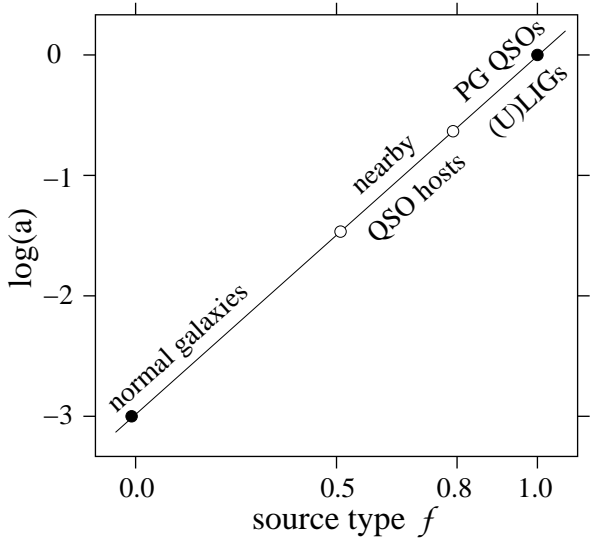}}
  \caption{Logarithm of the surface filling factor $a$ of star-forming gas as a function of source type $f$ as described in the text. The black filled circles mark the fits to the LIG/ULIG and the normal galaxies, respectively. The white filled circle indicate the region with which the nearby low-luminosity QSO hosts (discussed in this paper) are located.}
  \label{Figaf}
\end{figure}
Acknowledging the opaqueness of a dense, potentially continuous, medium and the importance of the contribution of individual star formation regions to the overall emission leads to the consequence that the sources are characterized by the surface filling factor $a$ of the star forming molecular gas in the host and a quantity $\beta$ that describes the $L_{\rm FIR}$ and $L'_{\rm CO}$ contributions of the emitting regions (see below).
Here $L'_{\rm CO}$  alone is not necessarily an adequate measure of the total 
molecular gas content in dense environments (as outlined in Sect. \ref{SectSFE}). Only if the standard assumptions are valid (virialized, i.e. gravitationally bound giant molecular clouds with different line-of-sight velocities), $I_{\rm CO}$ traces molecular gas in both the LIG/ULIG/PG QSO case  and the normal galaxy case. 

The two extreme populations (LIG/ULIGs/PG QSOs and normal galaxies) in Fig. \ref{FigLCOFIR} can then be described as follows. For PG QSO hosts and LIGs CO and FIR luminosity are directly proportional:
\
$$\log(L'_{\rm CO}) \varpropto \log(L_{\rm FIR}).$$
\
For normal galaxies, additional quiescent molecular gas reservoirs contribute to the object's total CO luminosity but do not, or at least not as efficiently, produce dust emission in the FIR. The additional contribution grows non-linear with increasing total molecular mass. 
The portion of CO-luminosity per FIR-luminosity is higher by a factor ($1 + \beta$),  with $\beta > 0$:
\
$$\log(L'_{\rm CO}) \varpropto [1+ \beta]  \log(L_{\rm FIR}) \sim  1.4  \log(L_{\rm FIR}). $$
\
As a first step we assume in the following a smooth transition between the host properties of these different classes, which can be parameterized by the value $f$ with 0$\le$$f$$\le$1.
\
$f$=1 represents the case of QSO hosts and LIGs/ULIGs, while $f$=0 represents the case of normal galaxies. 
\
Furthermore, a direct proportionality between the CO and FIR luminosity in the $f$=1 case and in the star forming portions of normal galaxies is assumed.
\
We also assume that in the $f$=1 case the molecular gas of a dominant part of the  host participates in the star formation process.  
\
In normal galaxies only a fraction $a$ - for which $\log(L'_{\rm CO}) \varpropto \log(L_{\rm FIR})$ is valid - participates in that process.
\
The quantity $a$ can be identified as the surface filling factor of the star forming molecular gas in the host with 0$<$$a$$<$1 (cf. Fig. \ref{FigActivity}).
\
The remaining part gives rise to $\beta  \log(L_{\rm FIR}$) as an additional portion of CO-luminosity. 
\
With a parameter $C_0$ we can then write:
\
\begin{eqnarray*}
\log(L'_{\rm CO}) &\cong& f  \log(L_{\rm FIR}) + \\
  & &(1-f)[\log(a L_{\rm FIR}) + \beta  \log((1-a)~L_{\rm FIR})] + C_0 \\
  \\
  &=&[1+(1-f)   \beta]  \log(L_{\rm FIR}) +\\ 
  & & (1-f)  [\log(a) + \beta  \log(1-a)] + C_0\\
\end{eqnarray*}
In Fig. \ref{Figaf} we show as a second step the interdependency of $a$ and $f$ as they result in suitable power-law fits to the extreme and intermediate source classes in Fig. \ref{FigLCOFIR}. We find that both quantities are linked via $$log(a) \cong 3(f-1).$$
\
\noindent
Within this model the two extreme source populations are represented by the following cases:
\\
Case I: for $f$=1, i.e. for QSO hosts and LIGs/ULIGs with $a$$\simeq$1 follows:
$$\log(L'_{\rm CO}) \cong  \log(L_{\rm FIR})  + C_0~.$$
With a value of $C_0$$\sim$-2, as suggested by Fig. \ref{FigLCOFIR}, this corresponds to the heuristically derived expressions for LIGs/ULIGs and PG QSOs
\citep{1991apj...370..158s, 1997apj...478..144s, 2003apj...585l.105s, 2006aj....132.2398e}.
\\
Case II: For normal galaxies with $f$=0, i.e. 0$<$$a$$\ll$1, and $\beta$$\sim$0.4
 the model yields:
$$\log(L'_{\rm CO}) \cong 1.4 \log(L_{\rm FIR}) - 5.0~.$$
This corresponds to the heuristically derived expressions for samples of distance limited or non-interacting, isolated FIR-selected galaxies \citep[][cf. Sect. \ref{SectSFE}]{1988apj...334..613s,1993a&a...272..123s}. 

We see that the formalism described above naturally results in the heuristic relations for the extreme source populations as presented in section 5.4 and shown in Fig. \ref{FigLCOFIR}.
It also allows us to describe the mean properties of nearby QSO hosts discussed in this paper as an intermediate population. The intermediate case of the nearby QSO hosts, for which 0.5$\lessapprox$$f$$\lessapprox$0.8 or $\log a$$\sim$-1, can then be approximated by
$$\log(L'_{\rm CO}) \cong (1.1 \pm 0.05) \log(L_{\rm FIR}) - (2.5 \pm 0.3)$$
or
$$L_{\rm FIR}/{10^9} \cong (50 \pm 20) \cdot [L'_{\rm CO}/{10^9}]^{0.91(\pm0.04)}~.$$

Overall the model corresponds to a picture in which all hosts are 
similar in the amount of molecular gas 
(i.e. about 10\% of $\sim 10^{12}$ M$_{\odot}$ of the total host mass) 
but different in the fraction of strong star forming regions. 
It also shows that the surface filling factor $a$ and the value $\beta$
that is linked to the FIR/CO emission properties of the source components 
are suitable parameters to describe a broad range of host galaxies.

\end{appendix}

\bibliographystyle{aa}
\bibliography{7578} 

\end{document}